\documentclass[prb,a4paper,twocolumn,showpacs]{revtex4}
\pdfoutput=1	
    \usepackage[utf8]{inputenc}
    \usepackage[T1]{fontenc}

    \usepackage{amssymb}
    \usepackage{amsmath}
    \usepackage{amsbsy}
    \usepackage{bm}
    \usepackage{graphicx}

    \newcommand{\I}{\mathrm{i}}
    \newcommand{\E}{\mathrm{e}}

\begin{document}
\title{Anisotropic dynamics of a vicinal surface under the meandering step instability}
\author{Alberto Verga}\email{Alberto.Verga@univ-provence.fr}
   \altaffiliation{%
    IM2NP, CNRS-UMR 6242, France.
   }
   \affiliation{%
   Aix-Marseille Université, IM2NP, 
   Campus de St Jérôme, Case 142, 13397 Marseille, France}

\date{\today}

\begin{abstract}
We investigate the nonlinear evolution of the Bales-Zangwill instability, responsible for the meandering of atomic steps on a growing vicinal surface. We develop an asymptotic method to derive, in the continuous limit, an evolution equation for the two-dimensional step flow. The dynamics of the crystal surface is greatly influenced by the anisotropy inherent to its geometry, and is characterized by the coarsening of undulations along the step direction and by the elastic relaxation in the mean slope direction. We demonstrate, using similarity arguments, that the coalescence of meanders and the step flow follow simple scaling laws, and deduce the exponents of the characteristic length scales and height amplitude. The relevance of these results to experiments is discussed. 
\end{abstract}


\maketitle

\section{Introduction}

The homoepitaxial growth of semiconductor and metallic vicinal surfaces, as revealed in experiments of molecular beam epitaxy, is characterized by a variety of morphological instabilities \cite{jeong99,politi00,Krug-2002cr,pierre-louis05}. The  surface evolves by step flow driven by the external flux and controlled by surface diffusion of adsorbed atoms, attachment rates (Schwoebel barriers) and elastic interactions. The interplay of different physical mechanisms can lead to the formation of characteristic structures at the nanometer scale, in the form for example of macroscopic bunches of steps, or periodic meanders of monatomic steps. These spontaneously nanopatterned surfaces can be useful as templates in an experimental two steps-process, to obtain by subsequent heteroepitaxy, self-organized arrays of quantum dots \cite{Ronda-2004ly,lichtenberger05,Berbezier-2007ve}. The physical description of the epitaxial surface growth depends on the characteristic lengths involved in the surface dynamics, from atomic processes (surface reconstruction and faceting) to gross macroscopic features (self-organization of nanostructures) \cite{shchukin99,stangl04}. At the mesoscopic scale of steps and terraces, the standard approach of Burton, Cabrera, and Frank \cite{burton51,pimpinelli98} completed with appropriated expressions for the equilibrium concentration and kinetic conditions, allows a fairly complete description of the vicinal surface. However, the macroscopic behavior of the vicinal surface, in particular the longtime nonlinear evolution of step bunches and meanders, can be more suitably accounted by a continuum model in terms, for instance, of partial differential equations for the surface height
\cite{Villain-1991zl,barabasi95}. 

In the continuum approach, initiated by Herring \cite{herring53} and Mullins \cite{mullins57}, the surface morphology is determined by the surface energy ($\gamma_S$, which is anisotropic in general) and the surface diffusion (characterized by the diffusion coefficient $D_S$). It is easy to incorporate into this model other thermodynamical contributions as, for instance, the elastic energy \cite{muller04}. To be more specific, let us consider the epitaxial growth of a crystal under a flux $F$ of atoms, relevant to molecular beam epitaxy experiments.\cite{barabasi95,pimpinelli98,saito98} We denote $a$ the height of an atomic layer. The front shape is given by the function $h(x,y,t)$ of the Cartesian coordinates $(x,y)$ and the time $t$. The thermodynamical state of the system is specified by a free energy functional $\mathcal{F}[h]$ of the crystal profile; the chemical potential per monolayer will be $\mu(h)=a\delta\mathcal{F}/\delta h$ ---as in the Cahn-Hilliard model.\cite{Cahn-1958ve} Therefore the conservative dynamics of the interface satisfies the Mullins equation \cite{mullins57}:
\begin{equation}
\frac{\partial}{\partial t}h=\frac{aD_S}{k_\mathrm{B}T}\nabla^2\mu(h)+a^3F\,,
\label{eq:mul}
\end{equation}
where $k_\mathrm{B}$ is the Boltzmann constant and $T$ the temperature. Equation (\ref{eq:mul}) is a mass conservation equation with the current $\bm{j}\sim-\nabla\mu(h)$. In the simplest case the chemical potential is proportional to the surface curvature $\mu=a^3\gamma_S\kappa$, and using a linear approximation, one obtains the evolution:
\begin{equation}
\frac{\partial}{\partial t}h=-B_S\nabla^4h+a^3F\,,
\label{eq:mul1}
\end{equation}
where we defined a mobility $B_S=a^4 D_S \gamma_S/k_\mathrm{B}T$. However, this approach, based on a free energy functional, valid for near-equilibrium conditions, prove to be insufficient when kinetic processes become important. Indeed, the attachment of adatoms to steps modifies the mobility coefficients (like $B_S$), and, coupled to external fluxes, can introduce new effects not related to a chemical potential. In particular, the appearance of step flow instabilities, bunching or meandering, cannot be described, in the continuum limit, by an evolution equation such as (\ref{eq:mul}) with $\mu$ derived from a variational functional. One example is the Bales-Zangwill instability \cite{bales90}, which in the weak nonlinear regime is described by the dimensionless equation \cite{frisch06,frisch07},
\begin{equation}
\frac{\partial}{\partial t}u=-\frac{\partial^2}{\partial y^2}\left[
 u + \frac{\partial^2}{\partial y^2}u + \left(
   \frac{\partial}{\partial y}u
   \right)^2
\right]\,,
\label{cks}
\end{equation}
where $u=u(y,t)$ is the rescaled step shape ($y$ is the step-wise direction). The first term on the right hand side, which is responsible for the instability, vanishes if the flux or the kinetic attachment barriers are absent. The third, nonlinear, term is also proportional to the flux and cannot be derived from a free energy functional. We investigate in this paper, the influence of the vicinal surface anisotropy on the two-dimensional dynamics of the meandering instability.

The meandering instability, first investigated by Bales and Zangwill  \cite{bales90}, results from the difference in the attachment kinetics of adatoms approaching a step in opposite directions, from the upper and lower terraces. The distinct neighborhood of an adatom sitting just above or at a step implies a difference in the height of the energy barriers that introduces a difference between the lower $\nu_+$ and upper $\nu_-$ attachment coefficients, the so-called Ehrlich-Schwoebel effect \cite{schwoebel66,ehrlich66}. Neglecting evaporation effects the instability growth rate, as mentioned above, is proportional to the flux and to the difference in kinetic coefficients $\sigma(q)\sim F(\nu_+-\nu_-)q^2$, where $q$ is the wavenumber of the perturbation. The original Bales-Zangwill linear analysis was extended to take into account various effects, such as desorption \cite{pimpinelli94}, kink barriers \cite{pierre-louis99,rusanen02PRB}, diffusion anisotropy \cite{frisch06} or elastic interactions \cite{paulin01,Yeon-2007rt}. Although experimental measures of the instability growth rate remain to be performed, there are detailed observations of step meandering in metals \cite{schwenger97,maroutian01,neel03}, and in semiconductors where it is often associated with step bunching, \cite{Tejedor-1998hc,schelling00,myslivecek02b,Omi-2000kx,Omi-2005ly} that reveal the longtime nonlinear stage of the instability. 

The longtime evolution of the Bales-Zangwill instability was thoroughly investigated in the strong nonlinear regime, by means of a continuous equation for the meander shape, derived from a multiscale analysis of the adatom diffusion model, that incorporates the effect of line step diffusion (related to kink barriers) \cite{pierre-louis98,kallunki00,gillet00,Hausser-2007ys} and the elastic interactions between steps \cite{paulin01} (in a one-dimensional approximation). This strong nonlinear model reproduces at least qualitatively, the meandering observed in metals \cite{maroutian99}, when desorption and nucleation can be neglected. The two-dimensional effects, together with mound formation were studied using kinetic Monte-Carlo simulations \cite{rost96,kallunki02,Kallunki-2004pd,sato05,nita05}.

In the case of semiconductors the coexistence of different kinetic instabilities leads to two-dimensional morphologies characterized by step bunches and meanders. In some circumstances, bunching and meandering is also present in metals \cite{neel03,Yu-2006mz}. Instability mechanisms operating in Si(001), based on the diffusion anisotropy that results from the surface reconstruction of alternating terraces, were found for bunching \cite{frisch05} and meandering \cite{frisch06}. One of the salient features of these two-dimensional patterns is their coarsening: roughening amplitude and length scale follow growing power laws. A unified continuum model of the bunching and meandering instability is lacking, even if phenomenological models can qualitatively account for some of its properties \cite{rost96,Krug-1999fj,Yu-2006mz}. The difficulty is to systematically derive, from the mesoscopic adatom diffusion model, a continuum limit that in addition to the kinetic processes, takes into account the step elastic interactions in its full two-dimensional form. In this paper we obtain such a model in the simplest physical situation, where only the meandering instability is present and local step interactions are considered. Under these assumptions we can develop a weak amplitude nonlinear expansion of the Burton, Cabrera and Frank equations. Our two-dimensional model of the meandering instability contrasts to previous ones which treated the nonlinear evolution of an in-phase pattern, thus reducing the problem to the (one-dimensional) dynamics of a single step. 

In the following section \S\ref{s:BCF} we present the basic equations, essentially the Burton, Cabrera and Frank model. Section \S\ref{s:eq} deals with the weak nonlinear expansion of the adatom diffusion equations, that allow us to derive in the continuum limit a differential equation for the surface height. In \S\ref{s:results} we study the evolution of the system by the numerical integration of the vicinal surface equation, and we derive using a self-similar solution, the asymptotic scaling laws for the amplitude and characteristic lengths. The last section \S\ref{s:conclusion} presents a discussion about a possible generalization of the model to include step bunching, and a concluding summary.

\section{Step flow: adatom diffusion equations}
\label{s:BCF}

\begin{figure}
\centering
\includegraphics[width=0.46\textwidth]{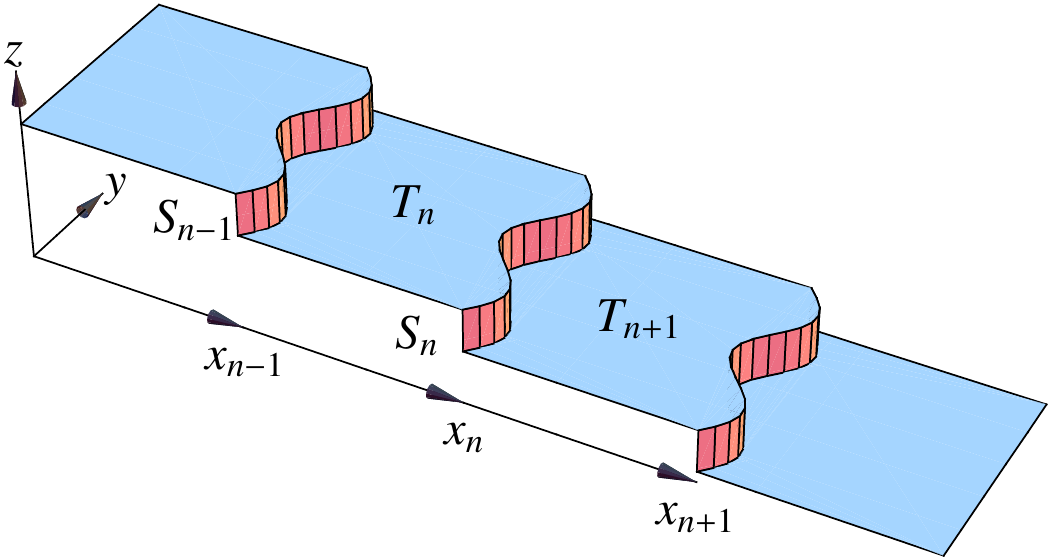}
\caption{Schematic geometry of the vicinal surface, formed by a set of terraces $T_n$ separated by atomic steps $S_n$ at $x=x_n(y,t)$.}
\label{fig:step}
\end{figure}

The geometry of the vicinal surface schematically represented in Fig.~\ref{fig:step}, can be described by the set of curves $x=x_n(y,t)$ representing the steps $n=0,1,\ldots,N$ at height $z=z_n=(N-n)a$. The terrace $T_n$ of initial size $l_0$, is bounded by the upper step $x_{n-1}$ and the lower step $x_n$; its slope is then $-m=-a/l_0$. We define the external normal and curvature of step $n$ by 
\begin{equation}\label{eq:nk}
	\bm{n}_n=\frac{(1,-\partial_yx_n)}{[1+(\partial_yx_n)^2]^{1/2}},\;
	\kappa_n=-\frac{\partial_{yy}x_n}{[1+(\partial_yx_n)^2]^{3/2}}\,,
\end{equation}
respectively. 

In the absence of an external flux $F$, there is an equilibrium concentration $C_{eq,n}$ of adatoms on the surface. This concentration, which is in general not uniform, depends on the curvature and elastic interactions of steps; it is determined by the surface chemical potential $\mu_n(y,t)=\mu_n^{(c)}+\mu_n^{(e)}$:
\begin{equation}\label{eq:ceq}
	C_{eq,n}=C_0\,\mathrm{e}^{\mu_n/k_\mathrm{B}T}\,,
\end{equation}
at temperature $T$. $C_0$ is the adatom concentration corresponding to the reference vicinal surface, the one consisting of equidistant straight steps. The first contribution $\mu_n^{(c)}$ to the chemical potential takes into account the step curvature, and is given by the Gibbs-Thomson relation, 
\begin{equation}\label{muc}
	\mu_n^{(c)}= \Omega\gamma_s \kappa_n\,,
\end{equation}
with $\gamma_s$ the step stiffness ($\Omega=a^2$ the atomic area). The second contribution $\mu_n^{(e)}$ results from the dipole moments created by the broken bonds at each step, and depends on the distance $x$ between steps (see for instance Marchenko \cite{marchenko80} or Duport et al. \cite{duport95}):
\begin{equation}\label{eq:mue}
	\mu_n^{(e)}=\Omega\beta_s(g_{n+1}-g_{n})\,,\;
	g_n=\frac{l_0^3}{(x_{n}-x_{n-1})^3}\,,
\end{equation}
where $\beta_s=4(1-\nu^2)m_s^2/\pi El_0^3$ has the dimensions of an (elastic) energy per unit surface, $\nu$ and $E$ are the Poisson ratio and the Young modulus respectively, and $m_s$ is the force dipole moment. We assume that step $n$ only interacts with its nearest neighbors; this approximation is justified in the case of the rapidly decreasing $1/x^3$ dipole interaction. In fact, summing over all steps would not change the form of the elastic repulsion in the continuum limit, but only renormalize the dimensional coupling constant $\beta_s$; as demonstrated by Xiang \cite{xiang02}, the ratio between the infinite series and one term of the sum (\ref{eq:mue}) is $\pi^2/6$. Under usual experimental conditions the energies associated to the step stiffness and the elastic repulsion are much smaller than the thermal energy, so the equilibrium concentration at step $n$ can be written as,
\begin{equation}\label{eq:Ceq}
    C_{eq,n}=C_0\left[1+\gamma l_0\kappa_n + \beta(g_{n+1}-g_{n})\right]\,,
\end{equation}
where we introduced the nondimensional parameters 
\begin{equation}\label{eq:GA}
	\gamma=\Omega\gamma_s/k_\mathrm{B}Tl_0\,,\;
	\beta=\Omega \beta_s/k_\mathrm{B}T\,.
\end{equation} 
It is worth noting that the ``equilibrium'' concentration changes with the geometry of the vicinal surface through the curvature of steps and their separation, it is used in fact as a reference to compute the supersaturation. 

In the presence of an external flux, the evolution of the adatom concentration $C_n=C_n(x,y,t)$ is well described by a quasi-stationary diffusion equation, as originally proposed in the model of Burton, Cabrera and Frank \cite{burton51},
\begin{equation}\label{eq:BCFdiff}
	D\nabla^2C_n(x,y,t)+F=0\,,
\end{equation}
where $D$ is the adatom diffusion coefficient, and the source term $F$ is the deposition flux (number of atoms per unit time and unit surface). In (\ref{eq:BCFdiff}) we neglected the time derivative by assuming a weak flux regime $F\ll D/\Omega l_0^2$. Under molecular beam epitaxy experimental conditions, with $F$ a few monolayers per minute, this `adiabatic' approximation of the diffusion equation is usually satisfied. Also the processes of evaporation and, eventually, nucleation of surface atoms are disregarded, their characteristic times being much more longer than the typical flow step time $1/Fl_0^2$. 

At steps, the boundaries of a terrace, adatoms are attached with a velocity rate $\nu_-$ if they come from the upper terrace, or $\nu_+$ if they come from the lower one. The Bales-Zangwill meandering instability \cite{bales90} appears in the case where $\nu_+>\nu_-$, the Ehrlich-Schwoebel effect \cite{ehrlich66,schwoebel66}. These parameters characterize the attachment kinetics that controls the flux of adatoms, then fixing their concentration at the terrace boundary:
\begin{equation}\label{eq:BCFbc1}
	D\,\bm{n}_{n-1}\cdot\nabla C_n=\nu_+(C_n-C_{eq,n-1}),\;x=x_{n-1}\,,
\end{equation}
for the upper step and,
\begin{equation}\label{eq:BCFbc2}
	D\,\bm{n}_{n}\cdot\nabla C_n=-\nu_-(C_n-C_{eq,n}),\;x=x_{n}\,,
\end{equation}
for the lower one. The step flow results from the balance between the diffusion fluxes (\ref{eq:BCFbc1}) and (\ref{eq:BCFbc2}) at each step. The normal velocity is given by
\begin{equation}\label{eq:BCFvn}
	V_n=\Omega D\,\bm{n}_n\cdot(\nabla C_{n+1}-\nabla C_{n}),\;x=x_n\,.
\end{equation}
Equations (\ref{eq:BCFdiff}), (\ref{eq:BCFbc1}), (\ref{eq:BCFbc2}) and (\ref{eq:BCFvn}) form a complete system from which we can determine the relevant physical parameters. It is convenient to introduce a reduced concentration,
\begin{equation}\label{eq:c}
	c=\frac{C-C_0}{C_0}\,,
\end{equation}
and nondimensional parameters related to the flux and the attachment coefficients,
\begin{equation}\label{eq:fa}
	f=\frac{Fl_0^2}{DC_0}\,,\;
	\alpha_\pm=\frac{\nu_\pm l_0}{D}\,,
\end{equation}
as well as to normalize lengths with $l_0$ (the terrace width) and time with $l_0^2/\Omega D C_0$, as derived from the normal velocity (\ref{eq:BCFvn}). Using these parameters we can write the set of equations in nondimensional form:
\begin{equation}\label{eq:diff}
	\nabla^2c_n(x,y,t)+f=0\,,
\end{equation}
\begin{multline}
	\bm{n}_{n-1}\cdot\nabla c_n=\alpha_+[c_n-\gamma \kappa_{n-1} - \beta(g_{n}-g_{n-1})]\,,\\
	\mathrm{at}\;x=x_{n-1}\,,
	\label{eq:bc1}
\end{multline}
\begin{multline}
	\bm{n}_n\cdot\nabla c_n=-\alpha_-[c_n-\gamma \kappa_n - \beta(g_{n+1}-g_{n})]\,,\\
	\mathrm{at}\;x=x_{n}\,,
	\label{eq:bc2}
\end{multline}
\begin{equation}	\label{eq:vn}
	V_n=\bm{n}_n\cdot(\nabla c_{n+1}-\nabla c_{n})\,,\;\mathrm{at}\;x=x_{n}\,.
\end{equation}
Equations (\ref{eq:diff}-\ref{eq:vn}) describe at a mesoscopic level the diffusion of adatoms on a terrace $n$ of a vicinal surface, driven by the external flux $f$ and controlled by the attachment kinetics $\alpha_{\pm}$, the stiffness $\gamma$, and the elasticity $\beta$ of the bounding steps.

\section{Vicinal surface equations in the continuum limit}
\label{s:eq}

A stationary train of straight steps, advancing at constant velocity $x_n=n+ft$ is an unstable solution of the step flow. A perturbation $\xi(y,n,t)$ of the step profile,
\begin{equation}
x_n=n+ft+\xi(y,n,t)\,,
\label{eq:xpert}
\end{equation}
induces a modification of the terrace adatom concentration in the form
\begin{equation}
c_n(x,y,t)=c_n(x+ft)+\phi(x,y,n,t)\,,
\label{eq:cpert}
\end{equation}
where the first term is the stationary parabolic concentration:
\begin{equation}
c_0(x)=-\frac{x^2 f}{2}+
	\frac{f(2+\alpha_-) (\alpha_+x+1)}
	{2 \left(\alpha_+ +\alpha_+ +\alpha_+\alpha_-\right)},\;
	-1\le x\le 0\,,
\end{equation}
solution of (\ref{eq:diff}) with the boundaries (\ref{eq:bc1}) and (\ref{eq:bc2}) in the case of straight steps, $\xi(y,n,t)=0$. Note that as the flux $f$ tends to zero, the step velocity and the concentration vanish.

We are interested in obtaining the growth rate of the meandering instability in the long wavelength, continuum limit. In this case it is sufficient to consider the in-phase mode with
\begin{equation}
\phi(x,y,n,t)= \phi(x,n,t)\E^{\I qy}\,,
\end{equation}
and
\begin{equation}
\xi(y,n,t)= \xi(t)\E^{\I(kn+qy)}
\end{equation}
where $\phi(x,n,t)$ and $\xi(t)$ are small amplitudes, and $(k,q)$ is the wavevector of the perturbation. The long wavelength limit is obtained by making a development in powers of the wavevector (up to the forth order). To compute the continuum limit we introduce the notation $\Delta n=(n+1)-n$ for the separation between two steps. In this limit both lengths, the steps separation $l_0$ and the step height $a$ tend to zero, but the slope $m=a/l_0$ must be kept constant. Therefore, we formally have,
\begin{equation}
x_{n\pm1}=n\pm\Delta n +\xi(y,n\pm\Delta n,t)\,,
\end{equation}
and assume that $\xi$ is a smooth function of $n$, in the limit $\Delta n\rightarrow0$. The elastic term at $x=x_n$ in the boundary conditions becomes
\begin{multline}
\frac{\Delta n^2}{[\Delta n+\xi(n+\Delta n)-\xi(n)]^3}-\\
\frac{\Delta n^2}{[\Delta n+\xi(n)-\xi(n-\Delta n)]^3}\,,
\end{multline}
and a similar expression for the boundary at $x=x_{n-1}$, where the numerator in $\Delta n^2$ ensures the correct limit when $\Delta n\rightarrow0$ (we omitted the dependence in $y$ and $t$). We know that the meandering instability is driven by the flux $f$ and the difference in attachment coefficient $\delta=\alpha_+-\alpha_-$. We replace (\ref{eq:xpert}) and (\ref{eq:cpert}) in the diffusion equations and retain the lowest order in $\Delta n$, $f$, and $\delta$ to get the real part of the dispersion relation:
\begin{equation}
\sigma=\frac{f \delta }{4 \alpha }q^2 -\gamma q^4 -\frac{3}{2} \alpha  \beta k^4 -\frac{1}{2} (\alpha  \gamma +6 \beta )k^2 q^2 \,,
\label{eq:re}
\end{equation}
and the imaginary part
\begin{equation}
\omega=\frac{f}{6}k^3 +\frac{f}{2 \alpha }k q^2 \,,
\label{eq:im}
\end{equation}
where $\alpha=\alpha_-$ and the amplitudes grow as $\phi,\,\xi\sim \E^{\sigma t-\I\omega t}$. The first term in (\ref{eq:re}) is the destabilizing one, it is proportional to the product $f\delta$. If the long wavelength limit is introduced by replacing $k,\, q\rightarrow \epsilon k,\, \epsilon q$, in order to keep all terms we must assume that, near the instability threshold, the parameters can be considered small such that:
\begin{equation}
f\rightarrow \epsilon f,\;\delta\rightarrow\epsilon^3\delta,\;
\gamma\rightarrow\epsilon^2\gamma,\;
\beta\rightarrow\epsilon^2\beta\,,
\label{eq:scale}
\end{equation}
where the last two relations guarantee the balance of the instability growth term with the stiffness and elastic relaxation terms. Using these relations all terms in $\sigma$ become order $\mathcal{O}(\epsilon^6)$, and order $\mathcal{O}(\epsilon^4)$ in $\omega$, suggesting the introduction of two time scales, $\epsilon^6 t$ (instability and relaxation) and $\epsilon^4 t$ (dispersive waves). 

To summarize, the weak amplitude expansion of the step shape must be of the form,
\begin{equation}
x_n(y,t)=n+\epsilon\xi(\epsilon y,n;\epsilon^6 t,\epsilon^4 t)\,,
\label{eq:xweak}
\end{equation}
and that of the concentration,
\begin{equation}
c_n(x,y,t)=\sum_{m}\epsilon^mc_m(x,\epsilon y,n;\epsilon^6 t,\epsilon^4 t)\,,
\label{eq:cweak}
\end{equation}
where the variable $x$ is itself a function of $\epsilon$ through the boundary conditions, and the scaling (\ref{eq:scale}) is assumed. The equations of the model, become in terms of the slow variables:
\begin{widetext}
\begin{equation}
\partial_{xx}c(x,y,n)+\epsilon^2\partial_{yy}c(x,y,n)+\epsilon f=0\,,
\label{eq:ec}
\end{equation}
with
\begin{multline}
	\frac{(1,-\epsilon^2\partial_y\xi(y,n-\Delta n)\cdot
		(\partial_x c(x,y,n),\epsilon\partial_y c(x,y,n))}
		{(1+\epsilon^4[\partial_y\xi(y,n-\Delta n)]^2)^{1/2}} 
		=(\alpha+\epsilon^3 \delta)\biggl[
		c(x,y,n)-\epsilon^2 \beta G_{\epsilon}(n-\Delta n)+\\
	\epsilon^5 \gamma\frac{\partial_{yy}\xi(y,n-\Delta n)}
		{(1+\epsilon^4[\partial_y\xi(y,n-\Delta n)]^2)^{3/4}}\biggr]
\label{eq:etm}
\end{multline}
at $x=n-\Delta n + \epsilon\xi(y,n-\Delta n)$,
and
\begin{equation}
	\frac{(1,-\epsilon^2\partial_y\xi(y,n)\cdot
		(\partial_x c(x,y,n),\epsilon\partial_y c(x,y,n))}
		{(1+\epsilon^4[\partial_y\xi(y,n)]^2)^{1/2}} 
		=-\alpha\biggl[
		c(x,y,n)-\epsilon^2 \beta G_{\epsilon}(n)+
	\epsilon^5 \gamma\frac{\partial_{yy}\xi(y,n)}
		{(1+\epsilon^4[\partial_y\xi(y,n)]^2)^{3/4}}\biggr]
\label{eq:etn}
\end{equation}
at $x=n+ \epsilon\xi(y,n)$, where
\begin{equation}
G_{\epsilon}(n)=\frac{\Delta n^2}{(\Delta n+\epsilon \xi(y,n+\Delta n)-\epsilon\xi(y,n))^3}-
\frac{\Delta n^2}{(\Delta n+\epsilon\xi(y,n)-\epsilon\xi(y,n-\Delta n))^3}\,,
\end{equation}
and the normal velocity,
\begin{equation}
V_n=\frac{(1,-\epsilon^2\partial_y\xi(y,n)}
		{(1+\epsilon^4[\partial_y\xi(y,n)]^2)^{1/2}} 
		\cdot
		(\partial_x ,\epsilon\partial_y)\big[
		c(x,y,n+\Delta n)-c(x,y,n)\big]\bigg|_{x=n+\epsilon\xi(y,n)}\,,
\label{eq:evn}
\end{equation}
\end{widetext}
where the implicit dependence on the time variable is understood (time derivatives appear only in the explicit expression of the normal velocity).
Using the expansion (\ref{eq:xweak}-\ref{eq:cweak}) together with the Taylor series in $\Delta n$ of $\xi(y,n\pm\Delta n)$, one may solve (\ref{eq:ec}) with the boundary conditions (\ref{eq:etm}-\ref{eq:etn}) to obtain a series in $\epsilon$ of the normal velocity (\ref{eq:evn}); a few terms of this series are:
\begin{align}
V_n=&\epsilon^2 f\left(
	\Delta n \,\partial_n\xi+\Delta n^2\frac{1}{6}\partial_{nnn}\xi\right)
		\nonumber\\
	& -\epsilon^3\Delta n^2\frac{3}{2}\alpha\beta\partial_{nnnn}\xi
		\nonumber\\
	& +\epsilon^4\Delta n^2\frac{f}{2\alpha}\partial_{nyy}\xi
		\nonumber\\
	& -\epsilon^5\Delta n \left(3\beta+
		\Delta n\frac{1}{2}\alpha\gamma\right)\partial_{nnyy}\xi
		\nonumber\\
	& -\epsilon^7\Delta n \,\gamma\partial_{yyyy}\xi
		\nonumber\\
	& -\epsilon^7\Delta n^2\left[\frac{f\delta}{4\alpha}\partial_{yy}\xi+
		\frac{1}{4\alpha}\partial_{yy}\left(\partial_y\xi\right)^2\right]+\mbox{h.o.t}\,,
\label{eq:vnpert}
\end{align}
where the higher order terms (h.o.t), terms having higher order derivatives or powers in the step shape amplitude $\xi(y,n)$, are shown to be irrelevant in the continuum limit. Note that in (\ref{eq:vnpert}) the derivatives of $\xi(y,n)$ with respect to $n$ are themselves functions of the small parameter $\epsilon$. This can be made explicit using the representation of the surface in terms of the height function $h(x,y,t)$ satisfying the compatibility condition,
\begin{equation}
	\mathcal{F}(n,y,t)=z_n+h(x_n(y,t),y,t)=0\,,
\label{eq:F}
\end{equation}
which signifies that $h=\mathrm{const.}$ at a height level $z_n=(N-n)a/l_0$, $x=x_n(y,t)=n+\epsilon\xi(y,n,t)$. In this representation the steps are the level contours of the surface $z=h$.

\begin{figure*}
\centering
\includegraphics[width=0.45\textwidth]{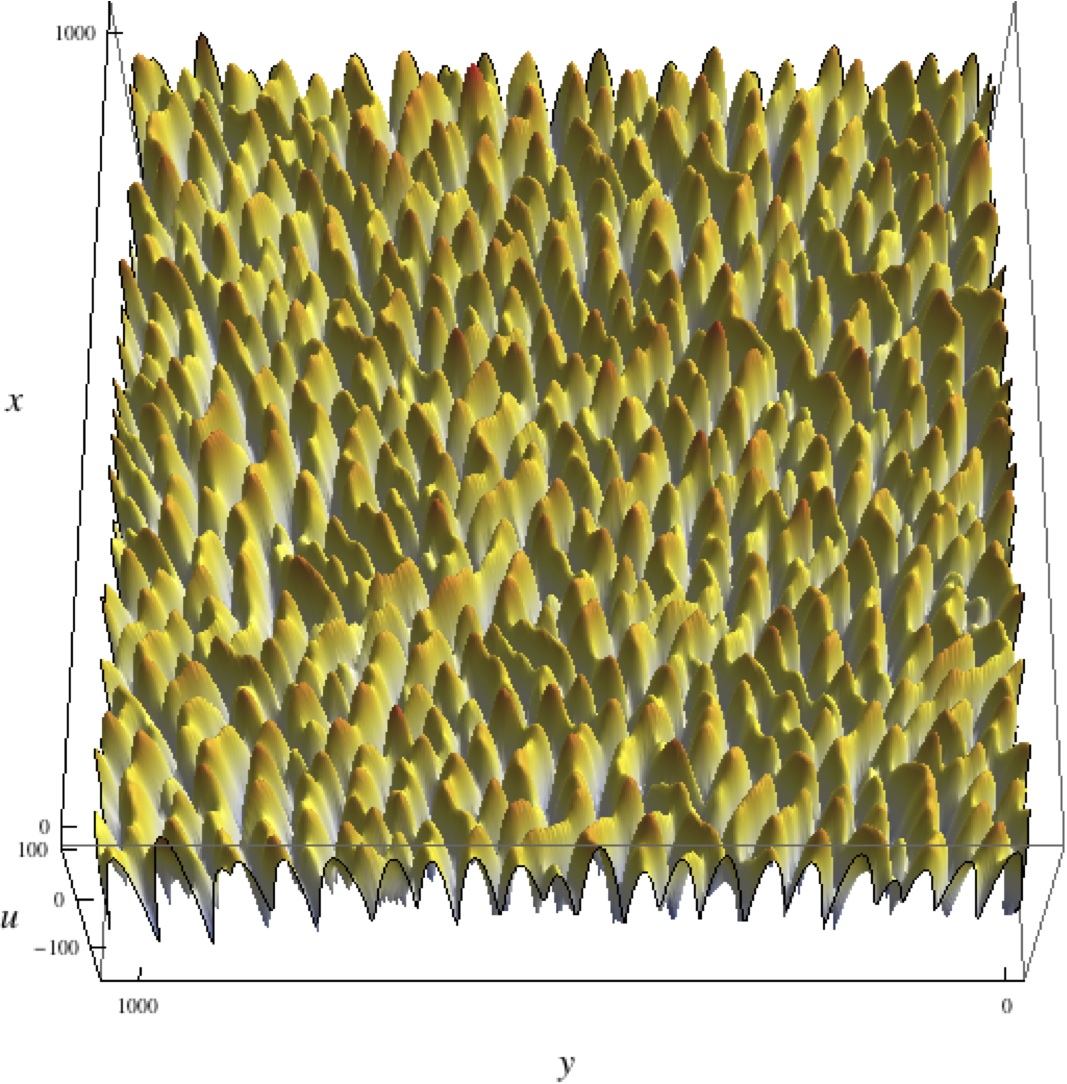}(a)%
\includegraphics[width=0.45\textwidth]{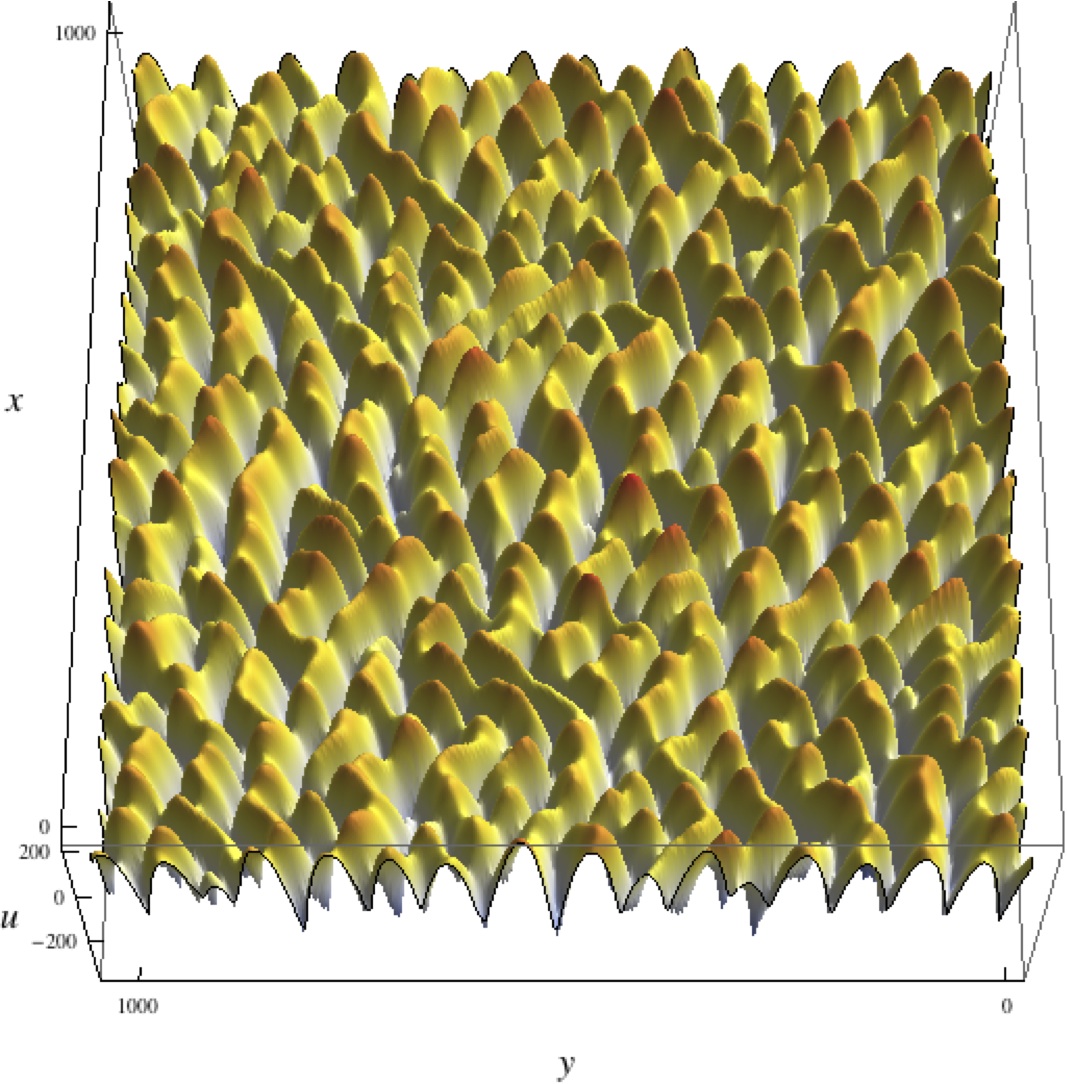}(b)\\
\includegraphics[width=0.45\textwidth]{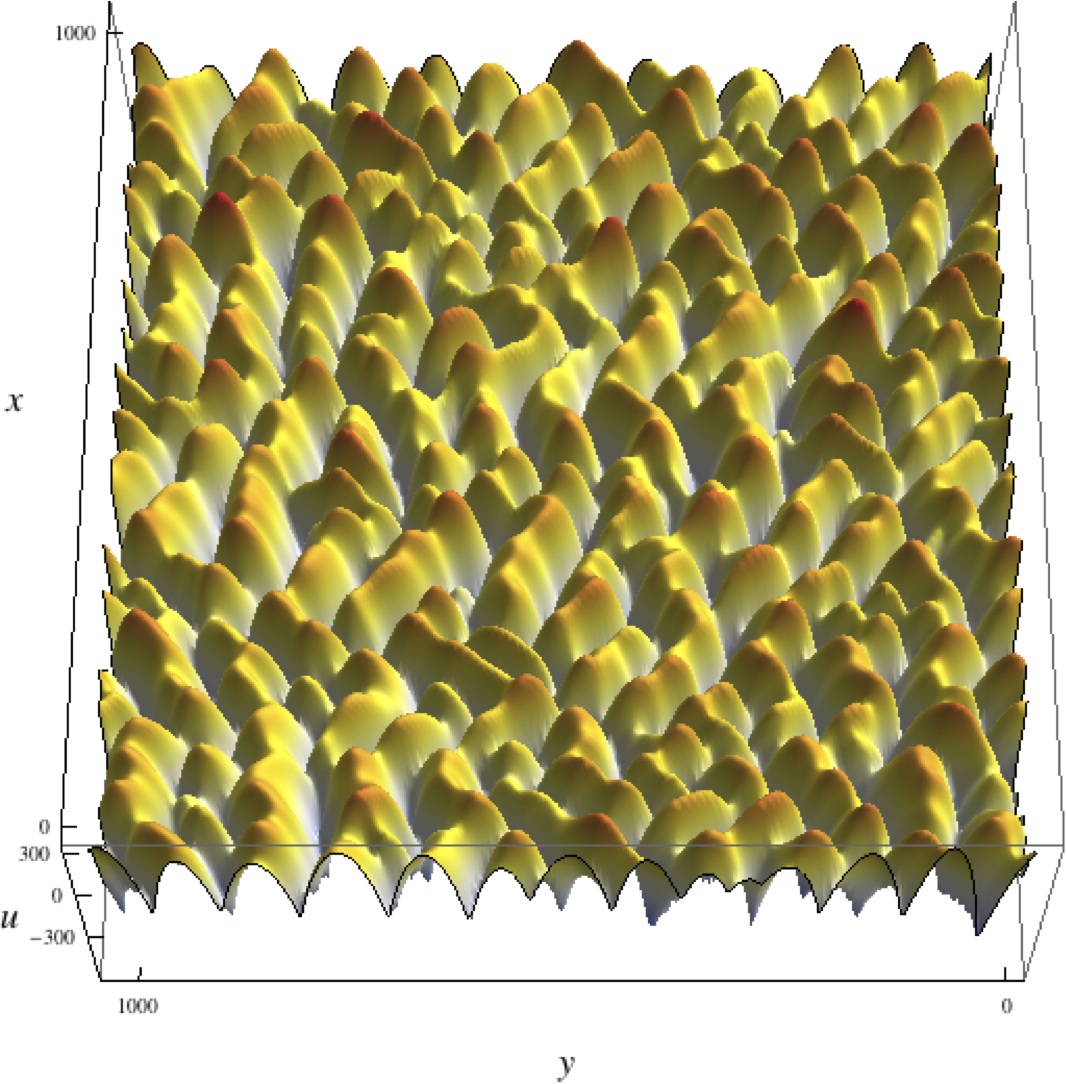}(c)%
\includegraphics[width=0.45\textwidth]{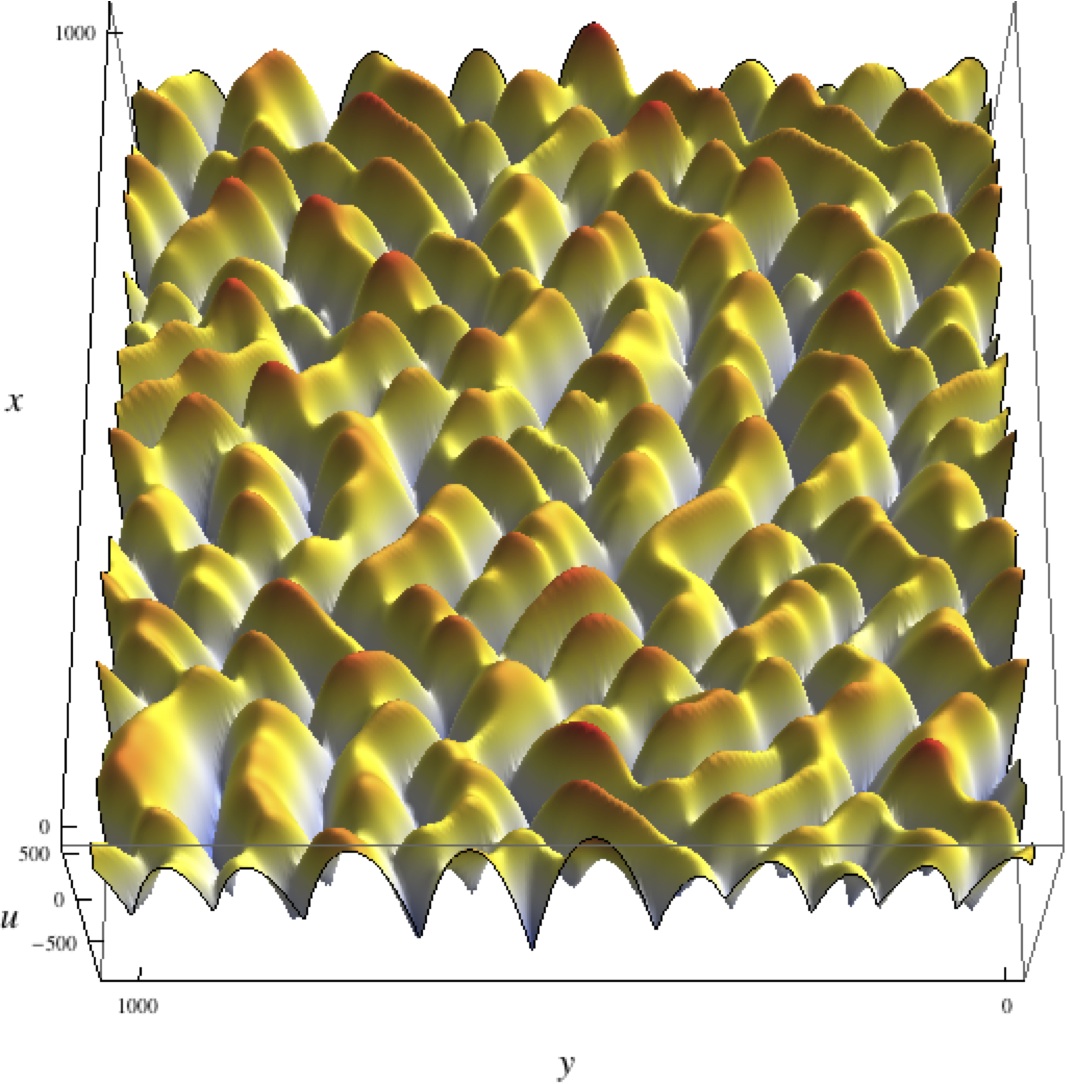}(d)
\caption{Surface height evolution, (a-d) $t=1000, 2000, 3000, 5350$. The system size is $1024^2$.}
\label{fig:h}
\end{figure*}

Using the derivatives of $\mathcal{F}(n,y,t)$ with respect to $n$ and $y$ one obtains the derivatives of $\xi$ in terms of derivatives of $h$. Typical derivatives are:
\begin{equation}
\partial_n x_n=\frac{m}{\partial_x h},\;
\partial_y x_n=-\frac{\partial_y h}{\partial_x h},\;
\partial_{nn} x_n=-\frac{m^2\partial_{xx}h}{(\partial_x h)^3}\,,
\end{equation}
etc. Moreover, the height itself deviates slightly from the mean slope $m$ (in the continuum longwave approximation) $h(x,y,t)=-m x+\epsilon u(\epsilon x,y,t)$. Keeping terms up to order $\epsilon^6$ (even number of derivatives) and $\epsilon^4$ (odd number of derivatives) that correspond to the two time scales, one obtains the normal velocity expressed in terms of the reduced height $u(x,y,t)$, with $(x,y,t)$ the slow variables and with $\Delta n=\epsilon=1$,\begin{multline}
	\partial_tu=-\partial_{xxxx}u-A\partial_{xxx}u\,-\\
	  \quad-\partial_{yy}\left[u+\partial_{yy}u+
	(\partial_yu)^2+B\partial_xu+C\partial_{xx}u\right] 
\label{eq:CKSa}
\end{multline}
where
\begin{align*}
	A=&\left(\frac{f}{\delta}\right)^{1/2}
			\left(\frac{2^3\gamma}{3^7\alpha\beta^3}\right)^{1/4}\,,\\
	B=&\left(\frac{f}{\delta}\right)^{1/2}
                        \left(\frac{2}{3\alpha^3\gamma\beta}\right)^{1/4}\,,\\
	C=&\left(\frac{6\beta}{\alpha\gamma}\right)^{1/2}+
		\left(\frac{\alpha\gamma}{6\beta}\right)^{1/2}\,,
\end{align*}
and where we used rescaled quantities in units of length $L_x$, $L_y$ (in both directions), time $T$ and amplitude $U$,
\begin{align*}
	L_x=&\frac{(24\alpha^3\gamma\beta)^{1/4}}{(f\delta)^{1/2}},\;
	L_y=\left(\frac{4\alpha \gamma}{f\delta}\right)^{1/2}\,,\\
	T=&16\gamma\left(\frac{\alpha}{f\delta}\right)^2,\;
	U=\frac{4m\alpha\gamma}{f}\,,
\end{align*}
and a moving frame with velocity $ft$ (this eliminates the first order derivative in $x$). Equation (\ref{eq:CKSa}) is the base of our numerical study of the morphological evolution of the vicinal surface under the Bales-Zangwill instability.

\section{Anisotropic coarsening of the vicinal surface}
\label{s:results}

\begin{figure}
\centering
\includegraphics[width=0.45\textwidth]{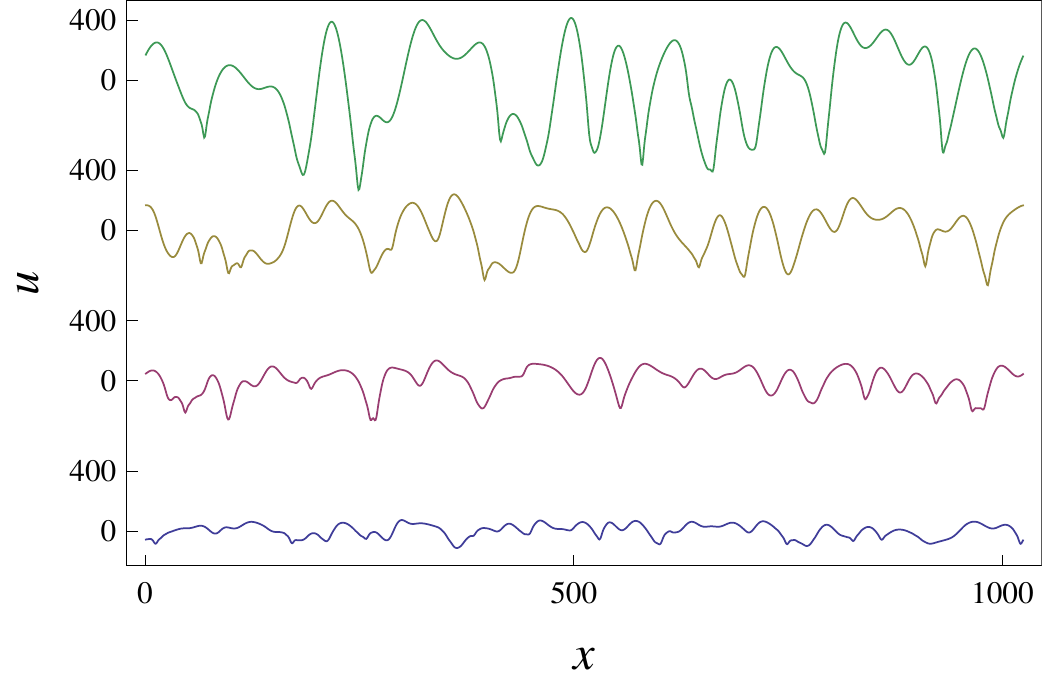}(a)\\
\includegraphics[width=0.45\textwidth]{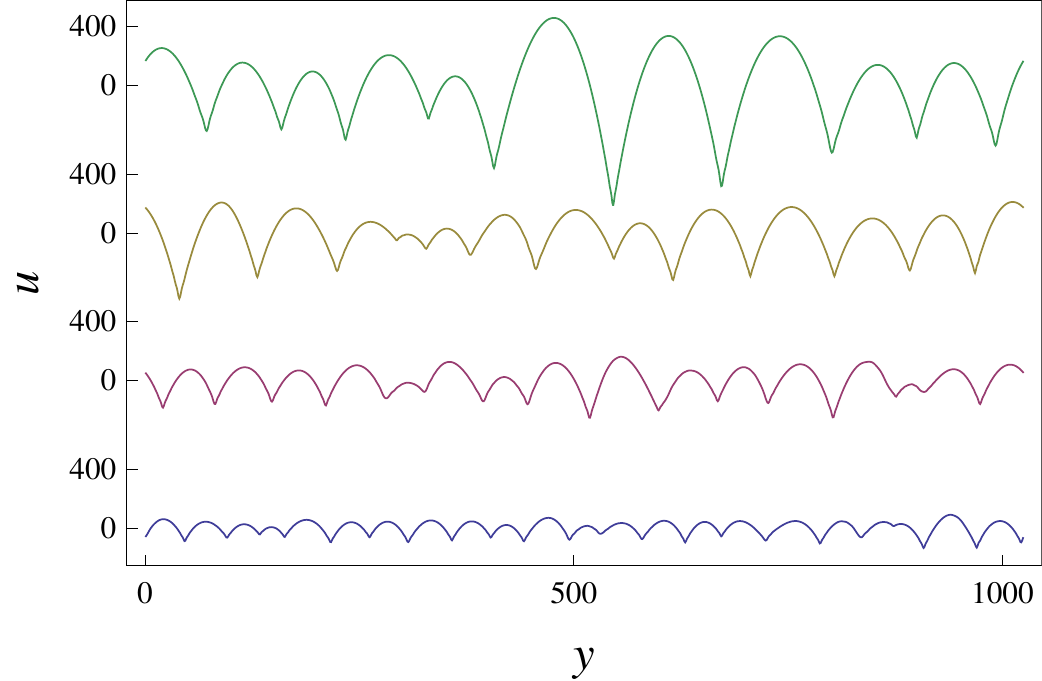}(b)
\caption{Coarsening dynamics in the $x$-direction (a) and in the $y$-direction (b). From bottom to top line-cuts of the height $u$ at $t=1000, 2000, 3000, 5350$.}
\label{fig:uxy}
\end{figure}

We integrate numerically (\ref{eq:CKSa}) using a pseudospectral method for space discretization and the exponential time differencing Runge-Kutta method for time stepping \cite{Kassam-2005yq}. Periodic boundary conditions are imposed. The typical space grid is $1024^2$ with $\Delta x=1\,l_0 L_x=1\,l_0 L_y$, and the time step $\Delta t=0.01\,(l_0^2/\Omega D C_0)\,T$. Resolution and size of the simulation are enough to obtain reliable statistical data. Throughout we use units $l_x=l_0 \,L_x$, $l_y=l_0 \,L_y$, $t_0=(l_0^2/\Omega D C_0)\,T$, and $u_0=l_0\,U$, for lengths, times and heights, respectively.

\begin{figure}
\centering
\includegraphics[width=0.46\textwidth]{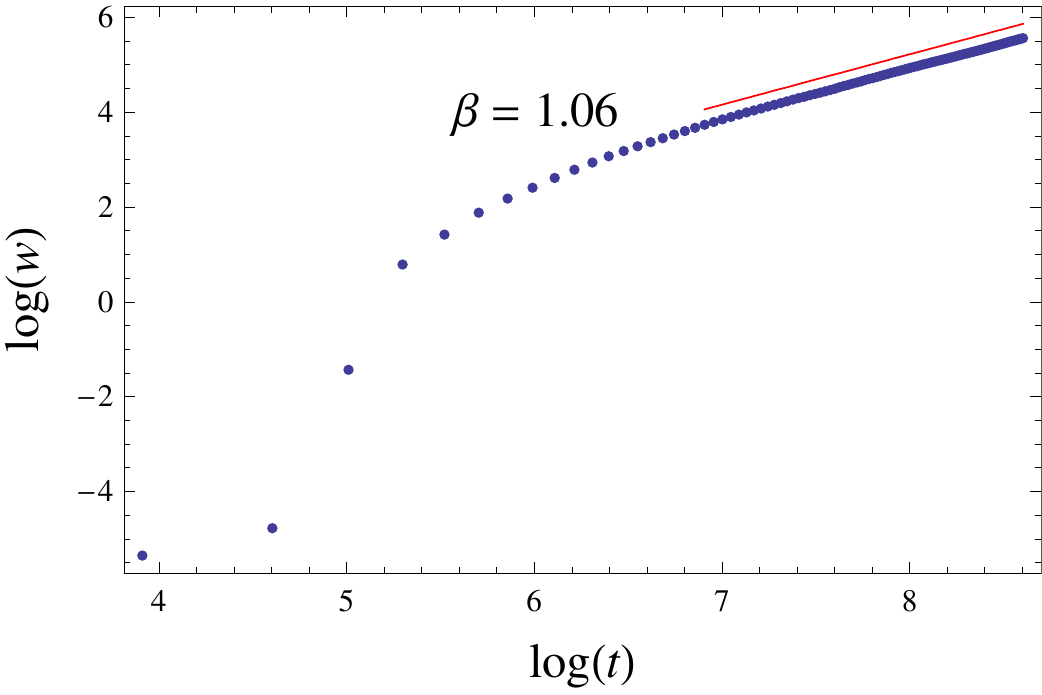}
\caption{Roughness as a function of time. Logarithmic plot showing the power law in $w\sim t$.}
\label{fig:w}
\end{figure}

We show the evolution of the vicinal surface at different times in Fig.~\ref{fig:h}. The horizontal axis is in the stepwise direction $y$, it gives the shape of the meanders, and the vertical one is in the terrace direction $x$, it follows downhill the mean slope, the third axis represent the fluctuations of the surface shape $u(x,y,t)$. We observe that the height scale, as well as the size of the structures steadily increase in time, which are characteristics of a coarsening dynamics. The surface develops in time an anisotropic pattern with parabolic meanders spanning in the $y$-direction together with complex fluctuations in the $x$-direction. In order to visualize the shape change of the surface in both directions we represent in Fig.~\ref{fig:uxy} a cut of the height (for example at the edge of the box) at fixed $y$ and $x$. Coarsening is observed along both directions, although the coarsening dynamics differs between step flow ($x$-direction) and meanders ($y$-direction). In the $y$-direction the meanders are composed by a series of parabola-like segments reminiscent to the evolution of the one dimensional meandering instability \cite{frisch07} (Fig.~\ref{fig:uxy}b). In particular, at late times, the system instead of tending towards a quasi-one-dimensional in-phase state with large meanders as would be dictated by the sole linear instability, a full two-dimensional state with persisting fluctuations remains (c.f. Fig.~\ref{fig:h}(d)). 

Using data from the time evolution one can compute the roughness 
$$
w(t)=\sqrt{\langle u(x,y,t)^2\rangle-\langle u\rangle^2}\,;
$$
(in our case the mean value $\langle u\rangle=0$ vanishes) it is represented in the graph of Fig.~\ref{fig:w}, in logarithmic scales. We can fit the long time roughness by a power law $w\sim t^\beta$ with exponent $\beta= 1.06$.

\begin{figure}
\centering
\includegraphics[width=0.46\textwidth]{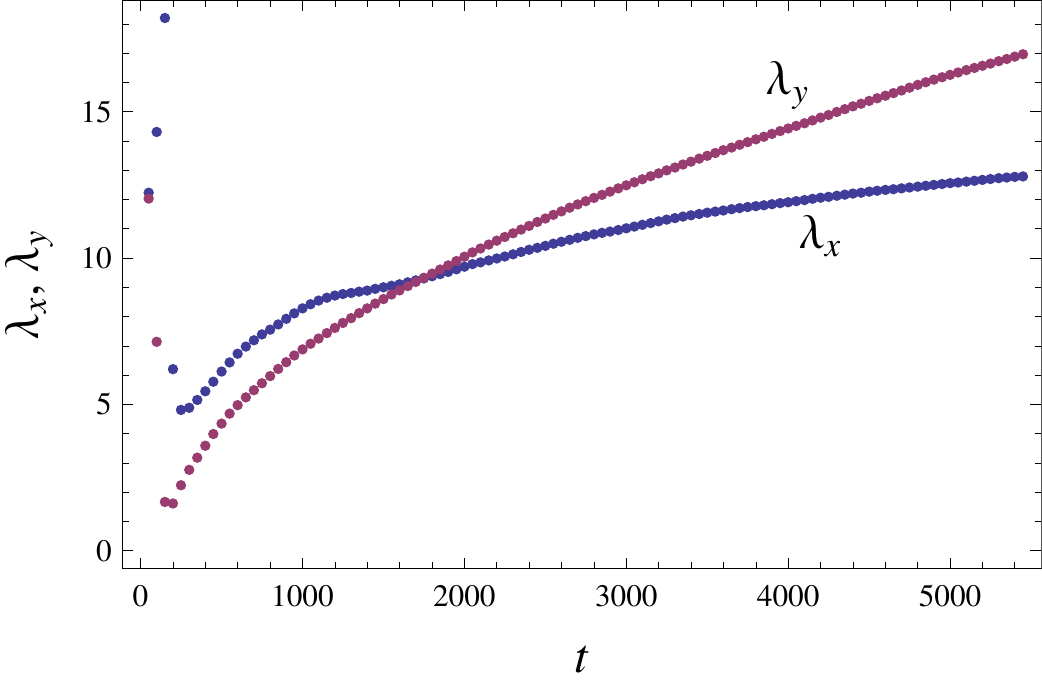}(a)\\[.7em]
\includegraphics[width=0.21\textwidth]{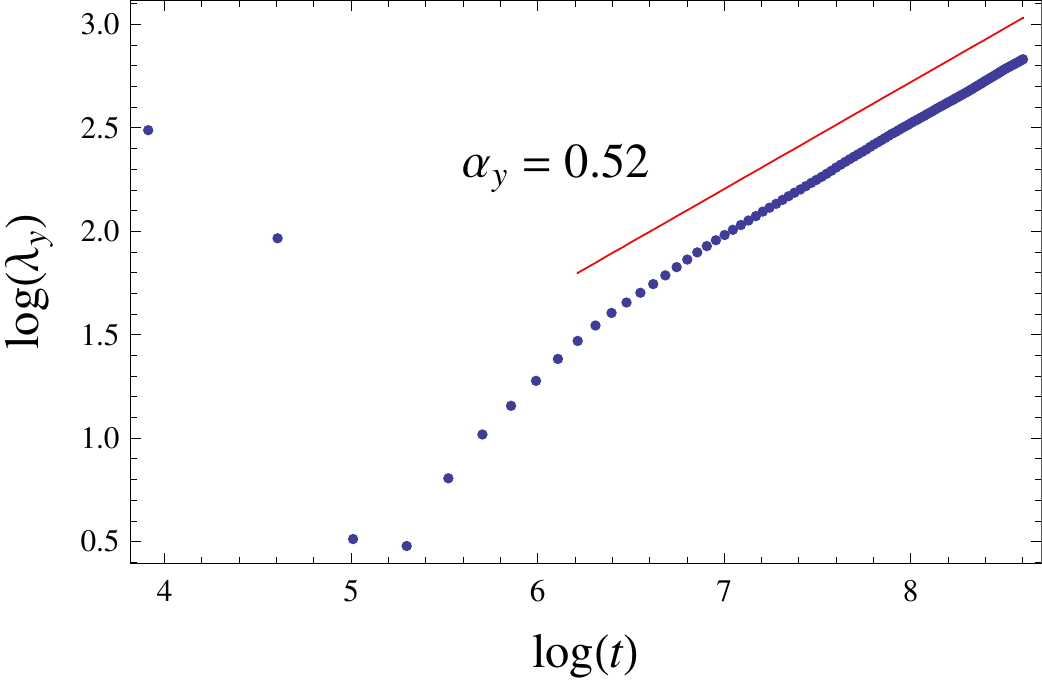}(b)\,%
\includegraphics[width=0.21\textwidth]{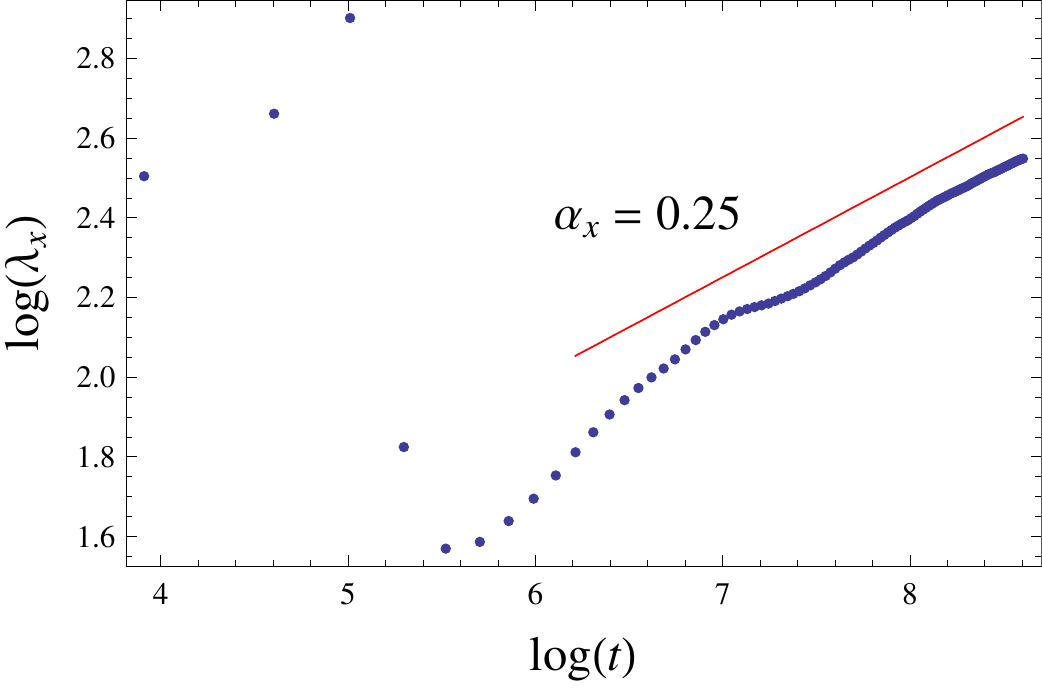}(c)
\caption{Coarsening lengths as a function of time. (a) The characteristic length increases faster in the meandering direction, $\lambda_y\sim t^{1/2}$ in red, and fit in (b), than in the step flow direction, $\lambda_x\sim t^{1/4}$ in blue, and fit in (c).}
\label{fig:lxly}
\end{figure}

\begin{figure}
\centering
\includegraphics[width=0.45\textwidth]{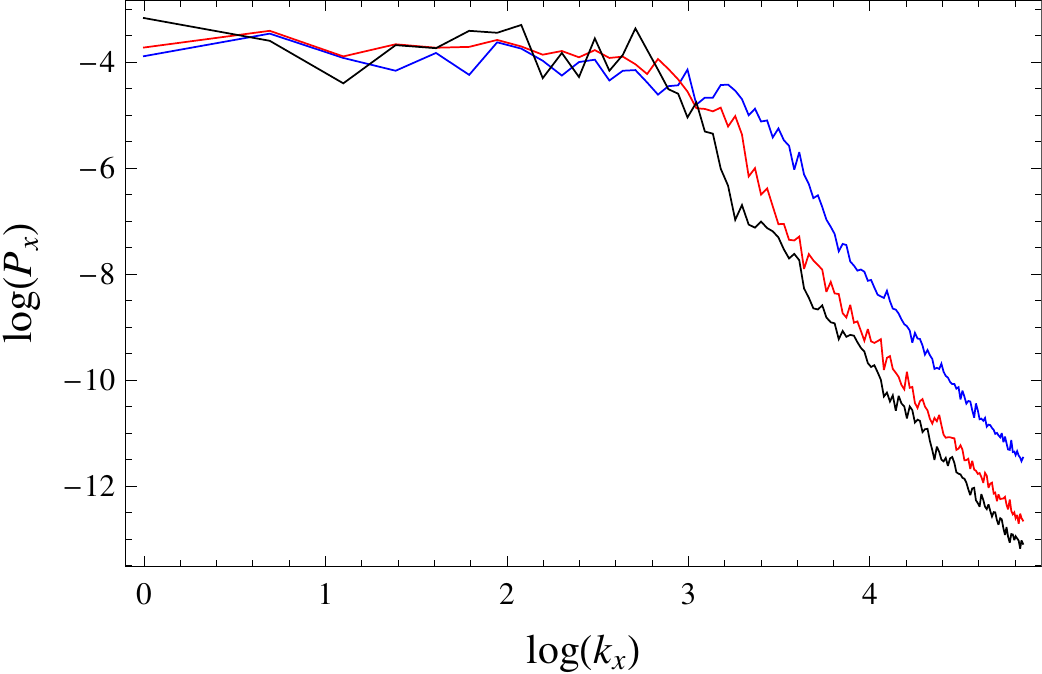}(a)\\
\includegraphics[width=0.45\textwidth]{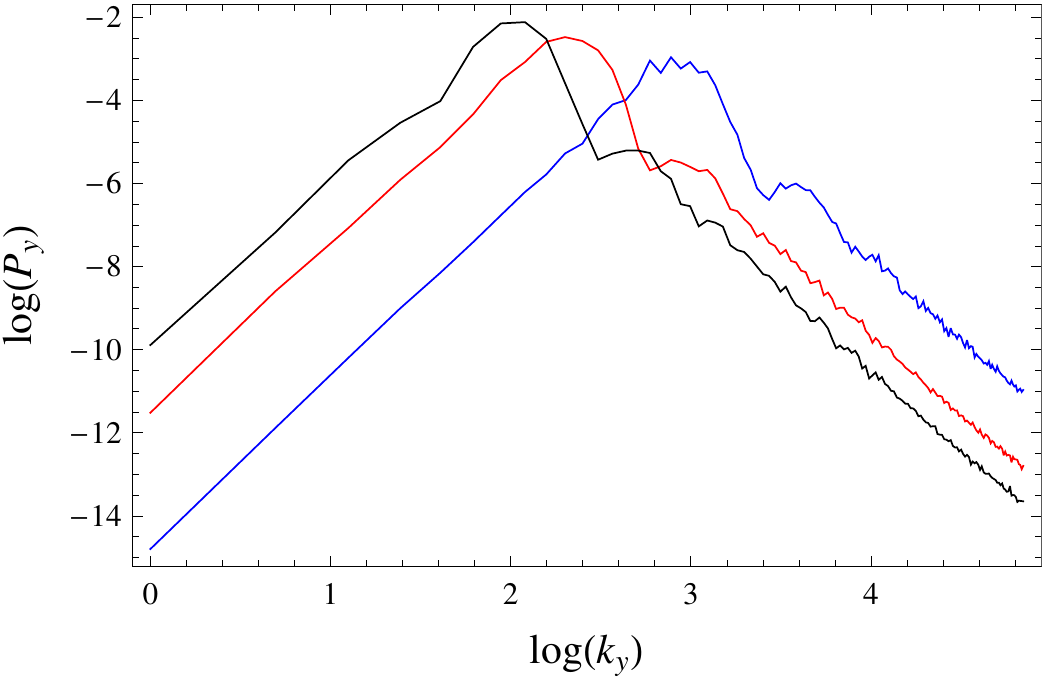}(b)
\caption{One-dimensional power spectrum (a) in the $x$ vicinal slope direction, and (b) in the $y$ step-wise direction. Three times are shown: 1000 (blue) 3000 (red) and 5350 (black).}
\label{fig:SPECn}
\end{figure}

In order to characterize the inherent anisotropic evolution of the surface morphology it is convenient to define characteristic lengths along both directions, perpendicular $\lambda_x$,
\begin{equation}
\lambda_x^{-2}=
\frac{\int dkdq \,k^2\,\left|u_{kq}\right|^2}{\int dkdq \left|u_{kq}\right|^2}\,,
\end{equation}
and parallel $\lambda_y$ to the steps,
\begin{equation}
\lambda_y^{-2}=
\frac{\int dkdq \,q^2\,\left|u_{kq}\right|^2}{\int dkdq \left|u_{kq}\right|^2}\,,
\end{equation}
where $u_{kq}(t)$ is the spatial Fourier transform of $u(x,y,t)$. Figure \ref{fig:lxly} shows the behavior of $\lambda_{x,y}(t)$ and the power law fits. At long times we obtain exponents of $\alpha_x=0.25$ and $\alpha_y=0.52$, for $\lambda_x$ and $\lambda_y$, respectively. This behavior can be correlated with the evolution of the power spectrum of the height fluctuations 
$$
P_x(k)=\frac{\int dq\, |u_{kq}(t)|^2}{\int dkdq\, |u_{kq}(t)|^2}\,,
$$ 
and
$$
P_y(q)=\frac{\int dk\, |u_{kq}(t)|^2}{\int dkdq\, |u_{kq}(t)|^2}\,,
$$ 
as shown in Fig.~\ref{fig:SPECn}. The spectrum of the terrace-wise fluctuations is rich at small wavevectors and shift towards the long wavelength direction in time; the stepwise fluctuations are peaked around a well defined wavelength (reminiscent of the initial most unstable mode), and as the coarsening develops the peak shift towards the long wavelength direction.

It is straightforward to explain the power laws observed in the numerical simulation of (\ref{eq:CKSa}); a simple similarity argument suffices. Equation (\ref{eq:CKSa}) satisfies the conservation relations,
\begin{equation}
\frac{d}{dt}\int dxdy\,u=0\,,
\end{equation}
the mean value of $u$ is constant, and
\begin{multline}
\frac{d}{dt}\int dxdy\,\frac{u^2}{2}=\int dxdy\,\big[(\partial_y u)^2-\\
	(\partial_{xx} u)^2-(\partial_{yy} u)^2-C(\partial_{xy} u)^2\big]
\label{eq:u2}
\end{multline}
the amplitude grows by the instability term preferentially in regions of strong gradients; therefore, as in the one-dimensional case \cite{frisch06}, the amplitude increase is controlled by the balance of the instability with the nonlinear term. On the other hand, we note that at long times the dispersive wave terms are dominated by the instability, relaxation and nonlinear terms. Indeed, linear terms give an increase in the characteristic wavelength $x,y\sim t^{1/3}$, faster than the relaxation length evolution $x,y\sim t^{1/4}$. Therefore, the effective long time equation contains essentially the driving instability term which competes with the nonlinear one, and the slowest relaxation term in the forth derivative of $x$:
\begin{equation}
\partial_t u=-\partial_{xxxx}u-\partial_{yy}\left[u+(\partial_y u)^2\right]\,.
\label{eq:effCKS}
\end{equation}
Putting the similarity ansatz,
\begin{equation}
u(x,y,t)=t^\beta U\left(\frac{x}{t^{\alpha_x}},\frac{y}{t^{\alpha_y}}\right)
\end{equation}
into (\ref{eq:effCKS}) one obtains the power law exponents $\beta=1$, and $\alpha_x=1/4$, $\alpha_y=1/2$, close to the numerical results (c.f Fig.~\ref{fig:lxly}). We confirm that this scaling is compatible with (\ref{eq:u2}), where the time derivative and the first two terms of the right hand side are of the same order, they increase linearly in time, while the two last terms rapidly become negligible ($(\partial_{yy} u)^2\sim \mathrm{const.}$ and $(\partial_{xy} u)^2\sim t^{1/2}$).

We note that the long time effective dynamics described by Eq.~(\ref{eq:effCKS}) is independent of the physical parameters. It accepts, as the original equation (\ref{eq:CKSa}), particular one-dimensional solutions $u=u(x,t)$ or $u=u(y,t)$. However, the equation is not separable (the product function $U(x,y)=U_x(x)U_y(y)$ is not a solution, because of the nonlinear term), showing that the long time behavior is essentially two-dimensional. The fact that the characteristic exponents of the meander coarsening are similar to the ones of the in-phase case, explains the reminiscence of the observed behavior to the one-dimensional case. In addition, as can be inferred from the partial spectra of Fig.~\ref{fig:SPECn}, the time dependence of the characteristic length scale of the step flow $\lambda_x$ is related, not to structures of $\lambda_x$ size, but to the steady shift of the spectrum towards small wavenumbers. This behavior contrasts with the $y$-direction spectrum, dominated by the characteristic size $\lambda_y$, of the parabolic meanders. Therefore, an appropriate description of the meander dynamics can be given by the coalescence of neighboring parabolic meanders, but at variance to the one-dimensional in-phase case, perturbed by the persistent step flow fluctuations. (It is worth mentioning, that adding a white noise term to (\ref{cks}) does not change its scaling behavior, as we found from direct numerical computations.)

\section{Conclusion}
\label{s:conclusion}

\begin{figure}
\centering
\includegraphics[width=0.45\textwidth]{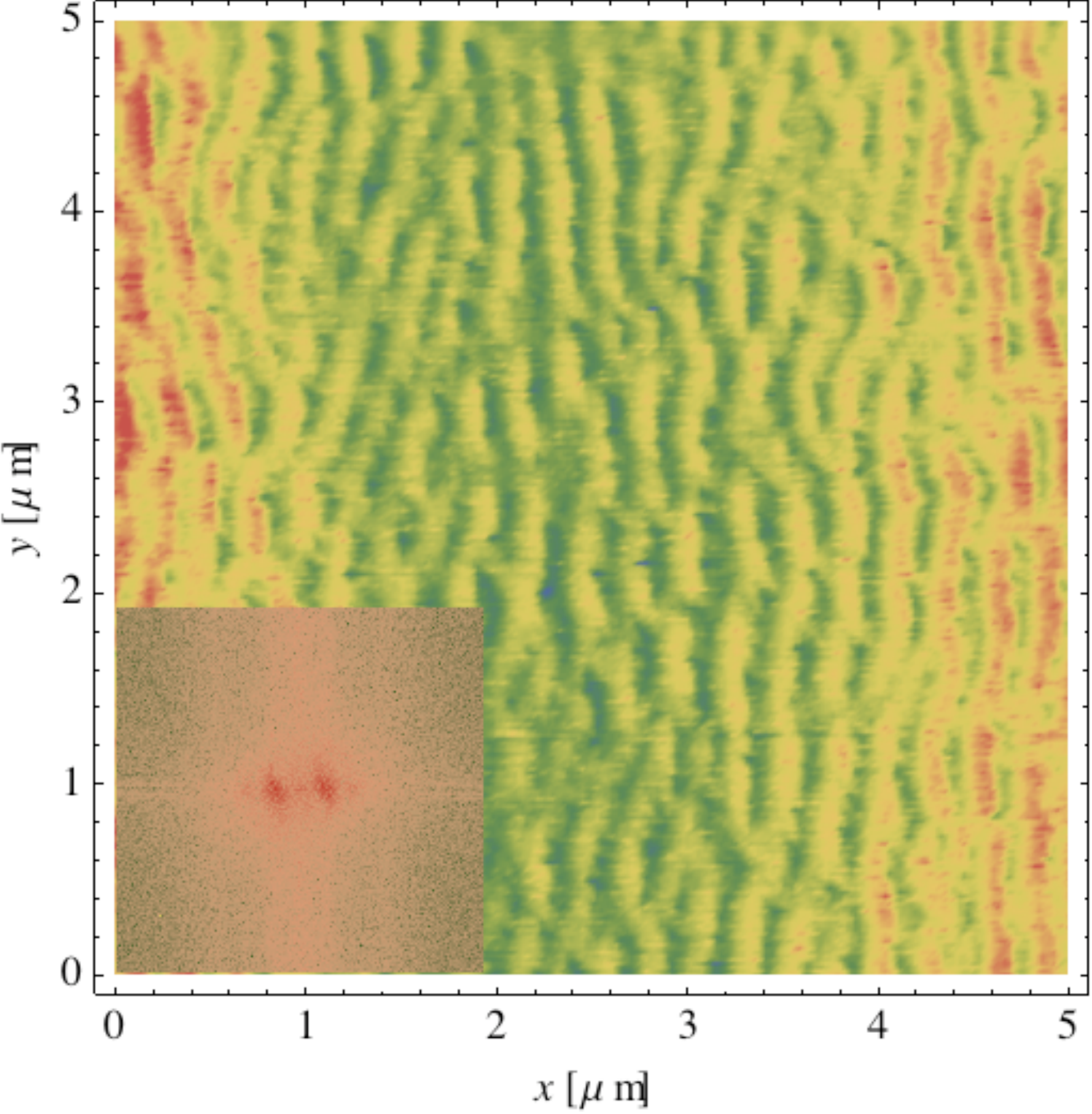}
\caption{Atomic force microscopy (AFM) image of Si(001) from experiments of Pascale et al. \protect{\cite{Pascale-2006kx}} after deposition of 300 monolayers. The system's size is $5\times5\,\mu\mathrm{m}^2$. Inset shows the Fourier power spectrum.}
\label{fig:AFM}
\end{figure}

In this paper we devised a method to obtain the evolution equation of a vicinal surface from the continuum limit of the adatoms diffusion microscopic model. We demonstrated that the anisotropy inherent to a vicinal surface induces different time scales for the amplitude and characteristic lengths parallel or perpendicular to the steps. As an interesting consequence an initial out of phase perturbation of the straight steps does not tend at long times towards the (most unstable) in phase pattern even if this symmetric mode is the most unstable one. Dephasing persists due to the slower coarsening in the step flow direction, $\lambda_x\sim t^{1/4}$, than the coarsening of meanders along the steps direction, $\lambda_y\sim t^{1/2}$. 

Although the present model can describe the weak nonlinear regime of the meandering instability under the condition that the step flow is stable, it would be important to generalize Eq.~(\ref{eq:CKSa}) in order to include the step bunching instability. The two instabilities can be present simultaneously even in homoepitaxy, in the case of anisotropic diffusion as in Si(001) \cite{frisch05,frisch06}. Experimental evidence of the persistence of two-dimensional patterns is presented in the atomic force image of Fig.~\ref{fig:AFM}, where we see a typical Si(001) vicinal surface grown by molecular beam epitaxy, after the deposition of 300 monolayers.\cite{Pascale-2006kx} In particular, at this late stage of the surface evolution, the form of the Fourier spectrum (inset) is somewhat similar to the one of Fig.~\ref{fig:SPECn}, but with the roles of $k_x$ and $k_y$ inverted. A possible form of the bunching-meandering instability equation is,
\begin{equation}
\partial_th=\hat{L}-\nabla^2|\nabla h|^2-\lambda\nabla\cdot\left(\partial_xh\nabla h\right)\,,
\label{eq:mb}
\end{equation}
where $\nabla=(\partial_x,\partial_y)$ and the linear operator $L$ is in Fourier space, of the form
\begin{equation}
\hat{L}=ak^2-k^4+bq^2-q^4+\mathrm{i}k(A k^2+ B q^2)- C k^2q^2\,,
\end{equation}
with $a,b$ constants that control the bunching and meandering instability growth rates, respectively, and  $\lambda,A,B.C=\mathrm{const.}$ depending on the physical parameters, in particular $\lambda$ must be proportional to the flux. Indeed, the nonlinear term is a generalization of the $F\partial_x(1/h_x)$ term, proportional to the flux, appearing in the unstable step flow. Derivation of (\ref{eq:mb}) from a microscopic model deserves further investigations. More generally the method introduced in this paper, valid in the weak amplitude, long wavelength approximation, applied here to the meandering instability, can be adapted to more general cases, notably to the case of heteroepitaxial growth of thin films on vicinal surfaces.

\acknowledgments
I would like to thank J.-N. Aqua, T. Frisch, and O. Pierre-Louis for valuable discussions; I also acknowledge collaboration with I. Berbezier and A. Ronda who provided me the data used in Fig.~\ref{fig:AFM} (ANR-PNANO MEMOIRE contract).

%\bibliography{steps}

\end{document}